\documentclass[12pt]{iopart}
\usepackage{lscape}
\usepackage{dcolumn}
\usepackage{iopams}
\usepackage{url}
\usepackage{graphicx}
\usepackage{xspace}

\newcommand{\eg}{\emph{e.g.}\ }
\newcommand{\ie}{\emph{i.e.}\ }

\newcommand{\beq}{\begin{equation}}
\newcommand{\eeq}{\end{equation}}
\newcommand{\beqa}{\begin{eqnarray}}
\newcommand{\eeqa}{\end{eqnarray}}
\newcommand{\beqn}{\begin{equation*}}
\newcommand{\eeqn}{\end{equation*}}
\newcommand{\beqan}{\begin{eqnarray*}}
\newcommand{\eeqan}{\end{eqnarray*}}

\newcommand{\ar}{\begin{array}}
\newcommand{\ear}{\end{array}}
\newcommand{\bc}{\begin{color}}
\newcommand{\ec}{\end{color}}
\newcommand{\bit}{\begin{itemize}}
\newcommand{\eit}{\end{itemize}}

\begin{document}
\title[\emph{Ab initio} study of the $^{33}$S $3p^4~^3P_J$ and $^{33}$S$^-$/~$^{37,35}$Cl $3p^5~^2P^o_J$ hyperfine structures.]{\emph{Ab initio} calculations of the $^{33}$S $3p^4~^3P_J$ and $^{33}$S$^-$/~$^{37,35}$Cl $3p^5~^2P^o_J$ hyperfine structures.}
\author{T Carette and M R Godefroid}

\address{Chimie Quantique et Photophysique, CP160/09, Universit\'e Libre de Bruxelles, \\ 
Av. F.D. Roosevelt 50, B-1050 Brussels, Belgium} 

\ead{tcarette@ulb.ac.be,mrgodef@ulb.ac.be}
\begin{abstract}
We present highly correlated multi-configuration Hartree-Fock (MCHF) calculations of the hyperfine structure of the $3p^5~^2P_J^o$ levels of $^{33}$S$^-$ and $^{35,37}$Cl. We obtain a good agreement with observation. The hyperfine structure of the neutral sulfur $^{33}$S $3p^4~^3P_J$ lowest multiplet that has never been measured to the knowledge of the authors, is also estimated theoretically.
We discuss some interesting observations made on the description of the atomic core in MCHF theory.
\end{abstract}

\submitto{\jpb}

\vfill
\noindent{\hfill \bf \today}

\maketitle

\section{Introduction}

As the experimental techniques evolve, the window of isotope effects in negative ions is becoming wider and wider~\cite{Traetal:89a,SunOls:90a,Beretal:95a,GodFro:99a,Bloetal:01a,Caretal:10a,CarGod:11a}. It offers new possibilities for understanding negative ions, correlation effects and the effect of nuclear properties on the electronic structure. In order to keep to this agenda, we need accurate models.

The hyperfine interaction, as arising mainly from the non-spherical nature of the electronic wave function near the nucleus, is an operator that is particularly sensitive to both core and valence correlation effects. It is also sensitive to contributions that do not affect the energy much.
When trying to calculate it, we face the difficulty of getting simultaneously a good description of the valence correlation -- crucial for negative ions --, core-valence and core correlation. It is therefore a useful tool for testing new approaches.

Experimental studies of the hyperfine interaction in atomic negative ions are scarce.
A pioneer study was made by Mader and Novick on the $^{3}$He$^-$ $1s2s2p~^4P^o$  metastable state~\cite{MadNov:74a}. 
Much more recently, the hyperfine structure of the only known E1 transition between bound states of a negative ion, the Os$^-$, was studied~\cite{Fisetal:10a}. In between, Trainham~\etal \cite{Traetal:89a} performed a study of the $^{33}$S$^-$ $3p^5~^2P^o_{3/2}$ hyperfine structure. The S$^-$ ion offers a good possibility for testing the methods on a middleweight system.

Recently, the isotope shift on the electron affinity of sulfur between the zero-spin isotopes 32 and 34 was measured and calculated~\cite{Caretal:10a}. A similar study involving the isotope 33 of sulfur, of spin $I=3/2$, would require the knowledge of the hyperfine structure of the involved states~\cite{Bloetal:01a}. However, if the $^{33}$S$^-$ ground state hyperfine structure is known, the one of the $^{33}$S $3p^4~^3P^{o}_{J}$ multiplet is not, to the knowledge of the authors.

We present in Section~\ref{sec5:S-Cl_HFS} large scale MCHF calculations on the iso-electronic systems S$^-$ and Cl, testing the possibility of using the valence correlation orbitals~\cite{Veretal:10a} for describing core effects. This choice of orbital set was successfully adopted for including a core-valence contribution in the calculation of the isotope shift on the electron affinity of sulfur~\cite{Caretal:10a} and chlorine~\cite{CarGod:11a}. In Section~\ref{sec5:S_HFS}, we calculate the hyperfine structure of the neutral sulfur lowest multiplet and point out a remarkable interplay of core and valence correlation in a MCHF description. 
We compare our theoretical results with experiment in Section~\ref{sec5:resHFS} and conclude in Section~\ref{sec5:concl} that some computational strategies provide a better definition for the core region.

\section{Theory}\label{sec5:th}

\subsection{The MCHF expansion}

The multiconfiguration Hartree-Fock (MCHF) variational approach is based on the following ansatz
\begin{equation}
\label{eq:SOC}
\Psi(\gamma LS M_L M_S) = \sum_{i} c_i \Phi(\gamma_i LSM_L M_S)
\end{equation} 
where the $\Phi(\gamma_i LSM_L M_S)$ are configuration state functions (CSF)~\cite{Froetal:97a} that are built on a single orthonormal basis set. In this theory, all $c_i$ are optimized and some -- if not all -- orbital radial functions are varied. We use the ATSP2K implementation of this method~\cite{Froetal:07a}. In this work, we first perform MCHF calculations to determine the orbital basis set. We then do subsequent larger configuration interaction (CI) calculations -- \ie optimizing only the $c_i$. An orbital active set is defined as the set of all orbitals characterized by quantum numbers with $n \leq n_{max}$ and $l\leq l_{max}$, and is denoted  $\lceil n_{max} l_{max} \rceil$.

Our scheme is based on the concept of ``multi-reference-interacting" (MR-I) set of configuration state functions detailed elsewhere~\cite{Caretal:10a}. It is defined as the set of all CSFs that interact to first order with at least one component of a given multi-reference. The $LS$ angular momenta of the occupied sub-shells are coupled by decreasing $n$ and $l$.

\subsection{Hyperfine interaction}\label{sec3:HFS}

The hyperfine structure of a $LSJ$ level is caused by the interaction of the electrons with the atomic nucleus of angular momentum quantum number $I$. The total atomic angular momentum quantum number is denoted $F$. To first order, the hyperfine energy correction has the form
\beqa
W(J,J) 
= A_J\frac{C}{2} + B_J\frac{3C(C+1)- 4 I(I+1)J(J+1)}{8I(2I-1)J(2J-1)}.
\eeqa
where $A_J$ and $B_J$ are respectively the magnetic dipole (M1) and electric quadrupole (E2) hyperfine constants and $C=F(F+1)-J(J+1)-I(I+1)$.
The theory underlying the computation of hyperfine structures using MCHF wave functions can be found in references~\cite{LinRos:74a,Hib:75a,Jonetal:93a}. 
It is possible to express the non relativistic hyperfine interaction in terms of the $J$-independent orbital~($a_{l}$),
spin-dipole~($a_{d}$), contact~($a_{c}$) and electric quadrupole ($b$) electronic hyperfine parameters~\cite{LinRos:74a}
\beqa
\label{eq3:small_a_l}
a_{l}&\equiv&\langle \Gamma LSM_{L}M_{S}|\sum_{i=1}^{N}
l^{(1)}_{0}(i)r^{-3}_{i}|\Gamma LSM_{L}M_{S}\rangle \; ,  \\
\label{eq3:small_a_sd}
a_{d}&\equiv&\langle \Gamma LSM_{L}M_{S}|\sum_{i=1}^{N}
2C^{(2)}_{0}(i)s^{(1)}_{0}(i)r^{-3}_{i}|\Gamma LSM_{L}M_{S}\rangle  \; , \\
\label{eq3:small_a_c}
a_{c}&\equiv&\langle \Gamma LSM_{L}M_{S}|\sum_{i=1}^{N}
2s^{(1)}_{0}(i)r^{-2}_{i}\delta(r_{i})|\Gamma LSM_{L}M_{S}\rangle  \; , \\
\label{eq3:small_b_hfs_par}
b &\equiv&\langle \Gamma LSM_{L}M_{S}|\sum_{i=1}^{N}
2C^{(2)}_{0}(i)r^{-3}_{i}|\Gamma LSM_{L}M_{S}\rangle  \; ,
\eeqa
and calculated for the magnetic component $M_{L}=L$ and $M_{S}=S$~\cite{Hib:75a}.

\subsection{The distinction between core and valence electrons}\label{sec2:CV}

When studying an atomic system, it is common to distinguish core and valence electrons based on the mono-configuration approximation.
This separation can usually be made quite efficiently by choosing a rare-gas-like core. In our case, the core is neon-like ($1s^22s^22p^6$) and the valence is composed of the $n=3$ electrons ($3s^23p^5$ for S$^-$ and Cl, and $3s^23p^4$ for S).
From that distinction, a lot of concepts based on a first-order picture of the correlation arise: core polarization, valence, core-valence and core-core correlation (V, CV, CC), etc. This terminology remains vague and the contributions those terms refer to often depend on the method~\cite{Veretal:10a,FroSax:74a,LinMor:86a,MigKim:98a}. In MCHF theory it is in general impossible to have a clean partition of those different contributions due to systematic rearrangements of the $\{c_i\}$ MCHF eigenvector and orbital shapes from one model to another.
The variational procedure is not perturbative and the core-valence distinction is based on a perturbative picture of the correlation. Still, the core-valence separation has a physical background so that it should not be discarded. Hence, we need a reliable definition of the core in order to interpret the so-called ``core effects".

In this work, the uncorrelated core is defined as being the $N_c$-electron core state built on the Hartree-Fock orbital set optimized  for the total system. In case of multiple solutions in the HF model, we select the one satisfying the Koopmans requirement~\cite{Fro:77a}.

\section{Hyperfine Structures of the $3p^5~^2P^o_J$ levels of S$^-$ and Cl}\label{sec5:S-Cl_HFS}

\subsection{Definition of the orbital set: MCHF calculations}

To explore the impact of the choice of the MCHF model defining the orbital set on the results of the open-core CI models, we use two references: the single main configuration, denoted SR, and the set of all valence single and double (SD)-excitations in the $n=3$ shell, denoted MR, \ie
\beq
\textrm{MR} = \{1,2\}^{10}\{3s,3p\}^6\{3\}^2.
\eeq
We then perform HF frozen-core MCHF calculations on the SR-I and MR-I sets obtained by activating all the electrons in orbital basis sets ranging from $\lceil 4f \rceil$ to $\lceil 13h \rceil$ and $\lceil 10h\rceil$ for SR and MR, respectively. The so-obtained one-electron radial functions are denoted SR-I-C and MR-I-C. In addition, we perform valence MCHF calculations on the \mbox{MR-I$\lceil nl \rceil$}, $n=4-13$, $l\leq 5$ ($h$ orbitals).  The resulting basis set is denoted MR-I-V. In all calculations, the [Ne] core is frozen to its HF shape.

Tables~\ref{tab5:weightsCl} and~\ref{tab5:weightsS-} display, respectively for Cl and S$^-$, the weights of the first configurations in the various MCHF calculations with $n_{max}l_{max}=10h$. The weight of a configuration is defined as
\beq
w= \left( \sum_i c_{i}^2 \right)^{1/2},
\eeq
where the sum runs over the CSFs belonging to the configuration.
We observe that the order of the few most important configurations is very similar in all cases. Still, some differences appear mainly due to the fact that some radial functions $P_{nl}(r)$ become inner orbitals in the MR-I-C and SR-I-C models. This is illustrated in the first four columns of Table~\ref{tab5:rad} in which we compare the mean radius of the spectroscopic and correlation orbitals of the MR-I-C model with the ones of the MR-I-V model for both Cl and S$^-$. In open-core calculations, most of the $n=5$ orbitals become inner orbitals and $P_{4d}, P_{4f}$ contract significantly. The SR-I-C orbitals do not differ strongly from the MR-I-C ones and are therefore not presented here.

\begin{table}
\caption{Sorted weights of the first configurations in the MR-I-V, MR-I-C and SR-I-C wave functions of Cl. The active set is $\lceil 10h \rceil$. The $1s$ and $2s$ sub-shells are closed in all those configurations.\label{tab5:weightsCl}}
\vspace{1mm}
\begin{indented}
\item[]
\begin{tabular}{cD{.}{.}{4}cD{.}{.}{4}cD{.}{.}{4}}
\br
\multicolumn{2}{c}{SR-I-C} &  \multicolumn{2}{c}{MR-I-C}  &\multicolumn{2}{c}{MR-I-V} \\
\crule{2}&\crule{2}&\crule{2}\\
 config. & \multicolumn{1}{c}{$w$} & config. & \multicolumn{1}{c}{$w$} & config. & \multicolumn{1}{c}{$w$}\\
\mr
 $2p^6 3s^2 3p^5          $ & 0.9712 & $2p^6 3s^2 3p^5          $ & 0.9584 &
$3s^2 3p^5$ & 0.9567 \\
 $2p^6 3s^2 3p^3 3d^2     $ & 0.1413 & $2p^6 3s^2 3p^3 3d^2     $ & 0.1877 &
$3s^2 3p^3 3d^2      $ & 0.1905 \\
 $2p^6 3s^1 3p^5 3d^1     $ & 0.0806 & $2p^6 3s^1 3p^5 3d^1     $ & 0.1121 &
$3s^1 3p^5 3d^1      $ & 0.1145 \\
 $2p^6 3s^2 3p^3 4p^2     $ & 0.0649 & $2p^6 3s^2 3p^3 4p^2     $ & 0.0653 &
$3s^2 3p^3 4p^2      $ & 0.0786 \\
 $2p^6 3s^1 3p^4 3d^1 4f^1$ & 0.0527 & $2p^6 3s^1 3p^4 3d^1 4f^1$ & 0.0538 &
$3s^1 3p^4 3d^1 4f^1 $ & 0.0691 \\
 $2p^6 3s^1 3p^4 3d^1 4p^1$ & 0.0411 & $2p^6      3p^5 3d^2     $ & 0.0416 &
$3s^1 3p^4 4s^1 4p^1 $ & 0.0584 \\
 $2p^6 3s^1 3p^4 4s^1 4p^1$ & 0.0403 & $2p^6 3s^1 3p^4 3d^1 4p^1$ & 0.0407 &
$3s^1 3p^4 3d^1 4p^1 $ & 0.0537 \\
 $2p^6      3p^5 3d^2     $ & 0.0346 & $2p^6 3s^1 3p^4 4s^1 4p^1$ & 0.0407 &
$3s^2 3p^4 4f^1      $ & 0.0475 \\
 $2p^6 3s^2 3p^4 4f^1     $ & 0.0311 & $2p^6 3s^2 3p^4 4f^1     $ & 0.0323 &
$     3p^5 3d^2      $ & 0.0417 \\
 $2p^6 3s^1 3p^4 4p^1 5s^1$ & 0.0310 & $2p^6 3s^1 3p^4 4p^1 5s^1$ & 0.0311 &
$3s^2 3p^3 4f^2      $ & 0.0370 \\
 $2p^6 3s^2 3p^3 3d^1 5g^1$ & 0.0269 & $2p^6 3s^2 3p^3 3d^1 5g^1$ & 0.0273 &
$3s^2 3p^4 4p^1      $ & 0.0322 \\
 $2p^4 3s^2 3p^5 4d^1 5d^1$ & 0.0268 & $2p^4 3s^2 3p^5 4d^1 5d^1$ & 0.0267 &
$3s^2 3p^3 3d^1 5g^1 $ & 0.0306 \\
 $2p^4 3s^2 3p^5 5d^2     $ & 0.0257 & $2p^4 3s^2 3p^5 5d^2     $ & 0.0253 &
$3s^2 3p^4 5p^1      $ & 0.0292 \\
 $2p^6 3s^2 3p^3 4f^2     $ & 0.0252 & $2p^6 3s^2 3p^3 4f^2     $ & 0.0248 &
$3s^1 3p^4 4p^1 4d^1 $ & 0.0271 \\
 $2p^4 3s^2 3p^5 5p^2     $ & 0.0249 & $2p^4 3s^2 3p^5 5p^2     $ & 0.0245 &
$3s^2 3p^3 4s^1 4d^1 $ & 0.0257 \\
\br
\end{tabular}
\end{indented}
\end{table}

\begin{table}
\caption{Sorted weights of the first configurations in the SR-I-C, MR-I-C and  MR-I-V wave functions of S$^-$. The active set is $\lceil 10h \rceil$. The $1s$ and $2s$ sub-shells are closed in all those configurations.\label{tab5:weightsS-}}
\vspace{1mm}
\begin{indented}
\item[]
\begin{tabular}{cD{.}{.}{4}cD{.}{.}{4}cD{.}{.}{4}}
\br
\multicolumn{2}{c}{SR-I-C} &  \multicolumn{2}{c}{MR-I-C}  &\multicolumn{2}{c}{MR-I-V} \\
\crule{2}&\crule{2}&\crule{2}\\
 config. & \multicolumn{1}{c}{$w$} & config. & \multicolumn{1}{c}{$w$} & config. & \multicolumn{1}{c}{$w$}\\
\mr
 $2p^6 3s^2 3p^5          $ & 0.9675 & $2p^6 3s^2 3p^5$           & 0.9522& %
$3s^2 3p^5$ & 0.9448 \\
 $2p^6 3s^2 3p^3 3d^2     $ & 0.1372 & $2p^6 3s^2 3p^3 3d^2     $ & 0.1931&%
$3s^2 3p^3 3d^2      $ & 0.1935 \\
 $2p^6 3s^2 3p^3 4p^2     $ & 0.0843 & $2p^6 3s^1 3p^5 3d^1     $ & 0.1124&%
$3s^1 3p^5 3d^1      $ & 0.1140 \\
 $2p^6 3s^1 3p^5 3d^1     $ & 0.0765 & $2p^6 3s^2 3p^3 4p^2     $ & 0.0848&%
$3s^2 3p^3 4p^2      $ & 0.1137 \\
 $2p^6 3s^1 3p^4 3d^1 4p^1$ & 0.0557 & $2p^6 3s^1 3p^4 3d^1 4p^1$ & 0.0550&%
$3s^1 3p^4 3d^1 4p^1 $ & 0.0755 \\
 $2p^6 3s^1 3p^4 4s^1 4p^1$ & 0.0496 & $2p^6 3s^1 3p^4 4s^1 4p^1$ & 0.0500&%
$3s^1 3p^4 4s^1 4p^1 $ & 0.0703 \\
 $2p^6 3s^1 3p^4 3d^1 4f^1$ & 0.0431 & $2p^6 3s^1 3p^4 3d^1 4f^1$ & 0.0441&%
$3s^1 3p^4 3d^1 4f^1 $ & 0.0624 \\
 $2p^6 3s^1 3p^4 3d^1 5f^1$ & 0.0424 & $2p^6 3s^1 3p^4 3d^1 5f^1$ & 0.0436&%
$3s^2 3p^4 4p^1      $ & 0.0606 \\
 $2p^6      3p^5 3d^2     $ & 0.0342 & $2p^6      3p^5 3d^2     $ & 0.0420&%
$3s^2 3p^4 4f^1      $ & 0.0494 \\
 $2p^6 3s^1 3p^4 4p^1 5s^1$ & 0.0328 & $2p^6 3s^1 3p^4 4p^1 5s^1$ & 0.0328&%
$3s^2 3p^4 5p^1      $ & 0.0463 \\
 $2p^4 3s^2 3p^5 5p^2     $ & 0.0309 & $2p^4 3s^2 3p^5 5p^2     $ & 0.0305&%
$     3p^5 3d^2      $ & 0.0418 \\
 $2p^4 3s^2 3p^5 4d^1 5d^1$ & 0.0296 & $2p^4 3s^2 3p^5 4d^1 5d^1$ & 0.0295&%
$3s^2 3p^3 4f^2      $ & 0.0355 \\
 $2p^6 3s^2 3p^3 3d^1 5g^1$ & 0.0276 & $2p^6 3s^2 3p^3 3d^1 5g^1$ & 0.0281&%
$3s^2 3p^3 4s^1 4d^1 $ & 0.0343 \\
 $2p^4 3s^2 3p^5 5d^2     $ & 0.0263 & $2p^6 3s^2 3p^4 5f^1     $ & 0.0272&%
$3s^1 3p^4 4p^1 4d^1 $ & 0.0342 \\
 $2p^6 3s^2 3p^4 5f^1     $ & 0.0260 & $2p^4 3s^2 3p^5 5d^2     $ & 0.0259&%
$3s^1 3p^4 3d^1 5f^1 $ & 0.0335 \\
\br
\end{tabular}
\end{indented}
\end{table}

\begin{table}
\caption{Mean radius (in $a_0$) of the $n=1$ to $5$ orbitals of Cl, S$^-$ and S in the \mbox{MR-I-V$\lceil 10h\rceil$} (valence) and \mbox{MR-I-C$\lceil 10h\rceil$} (open-core) models. $1s,2s,2p$ are Hartree-Fock orbitals.\label{tab5:rad}}
\begin{indented}
\item[]
\begin{tabular}{cD{.}{.}{5}D{.}{.}{5}D{.}{.}{2}D{.}{.}{5}D{.}{.}{5}D{.}{.}{2}D{.}{.}{5}D{.}{.}{5}}
\br
& \multicolumn{2}{c}{Cl} && \multicolumn{2}{c}{S$^-$} &&\multicolumn{2}{c}{S} \\
&\crule{2}&&\crule{2}&&\crule{2}\\
$nl$ & \multicolumn{1}{c}{valence} & \multicolumn{1}{c}{open-core} && \multicolumn{1}{c}{valence} & \multicolumn{1}{c}{open-core} && \multicolumn{1}{c}{valence} & \multicolumn{1}{c}{open-core} \\
\mr
$1s$ & 0.09130 & 0.09130 && 0.09715 & 0.09715 && 0.09715 & 0.09715\\
$2s$ & 0.44171 & 0.44171Ê&& 0.47585 & 0.47585 && 0.47577 & 0.47577\\
$2p$ & 0.40572 & 0.40572 && 0.44106 & 0.44106 && 0.44104 & 0.44104\\\\
$3s$ & 1.54735 & 1.55045 && 1.76214 & 1.76836 && 1.71095 & 1.71268\\
$3p$ & 1.82220 & 1.82778 && 2.28135 & 2.29572 && 2.02884 & 2.03898\\
$3d$ & 1.76250 & 1.75997 && 2.15941 & 2.16075 && 1.95878 & 1.96343\\
$4s$ & 2.03477 & 1.92840 && 2.44030 & 2.39304 && 2.40713 & 0.71754\\
$4p$ & 2.41053 & 2.13745 && 3.34104 & 2.80812 && 2.81064 & 0.70561\\
$4d$ & 2.23120 & 1.35932 && 2.86818 & 1.64193 && 2.45573 & 1.42118\\
$4f$ & 1.82616 & 1.30900 && 3.03237 & 1.17530 && 2.98537 & 1.20579\\
$5s$ & 2.14697 & 0.69456 && 2.71559 & 0.74676 && 2.28372 & 2.13212\\
$5p$ & 2.16166 & 0.68502 && 2.88173 & 0.75013 && 2.42314 & 2.38464\\
$5d$ & 2.08468 & 0.54938 && 2.04493 & 0.76611 && 1.70728 & 0.60688\\
$5f$ & 2.06913 & 0.65051 && 2.31986 & 1.55795 && 2.31174 & 0.63876\\
$5g$ & 1.85751 & 1.78797 && 3.40818 & 2.12699 && 3.83176 & 1.94307\\
\br
\end{tabular}
\end{indented}
\end{table}

\subsection{Fully correlated CI calculations}

The results of Tables~\ref{tab5:weightsCl} and~\ref{tab5:weightsS-} indicate that, even if the variational contents of open- and closed-core calculations are \emph{a priori} very different~\cite{Godetal:98a}, it is here possible to find extensions of the models developed above that remain comparable. Indeed, if we choose as multi-references the five or nine first configurations in the sorted lists
\beqa
\textrm{MR}_5&=&\{3s^23p^5, 3s^23p^33d^2,3s^13p^53d^1,3s^23p^34p^2,3s^13p^43d^14f^1\}\\
\textrm{MR}_9&=&\textrm{MR}_5\cup \{3s^13p^44s^14p^1, 3s^13p^43d^14p^1, 3p^53d^2, 3s^23p^44f^1 \}
\eeqa
for both systems, the selected MRs account for approximately the same correlation effects in all S$^-$ and Cl orbital sets. It allows a significant comparison between the CI calculations performed on the MR$_5$- and MR$_9$-I-C multi-reference-interacting sets performed with the orbital sets arising from  either the SR-I-C, MR-I-C or MR-I-V MCHF calculations.

\begin{figure}
\caption{For both S$^-$ (left plots) and Cl (right plots), values of non-relativistic magnetic dipole hyperfine interaction parameters $a_l, a_d$ and $a_c$ obtained by CI calculations in the MR$_5$-I-C model, in atomic units ($a_0^{-3}$), with the three explored orbital basis sets SR-I-C, MR-I-C and MR-I-V, as a function of the active space $\lceil nl\rceil$, $l\leq 5$. \label{fig5:conva}}
\center
\includegraphics[width=0.47\textwidth]{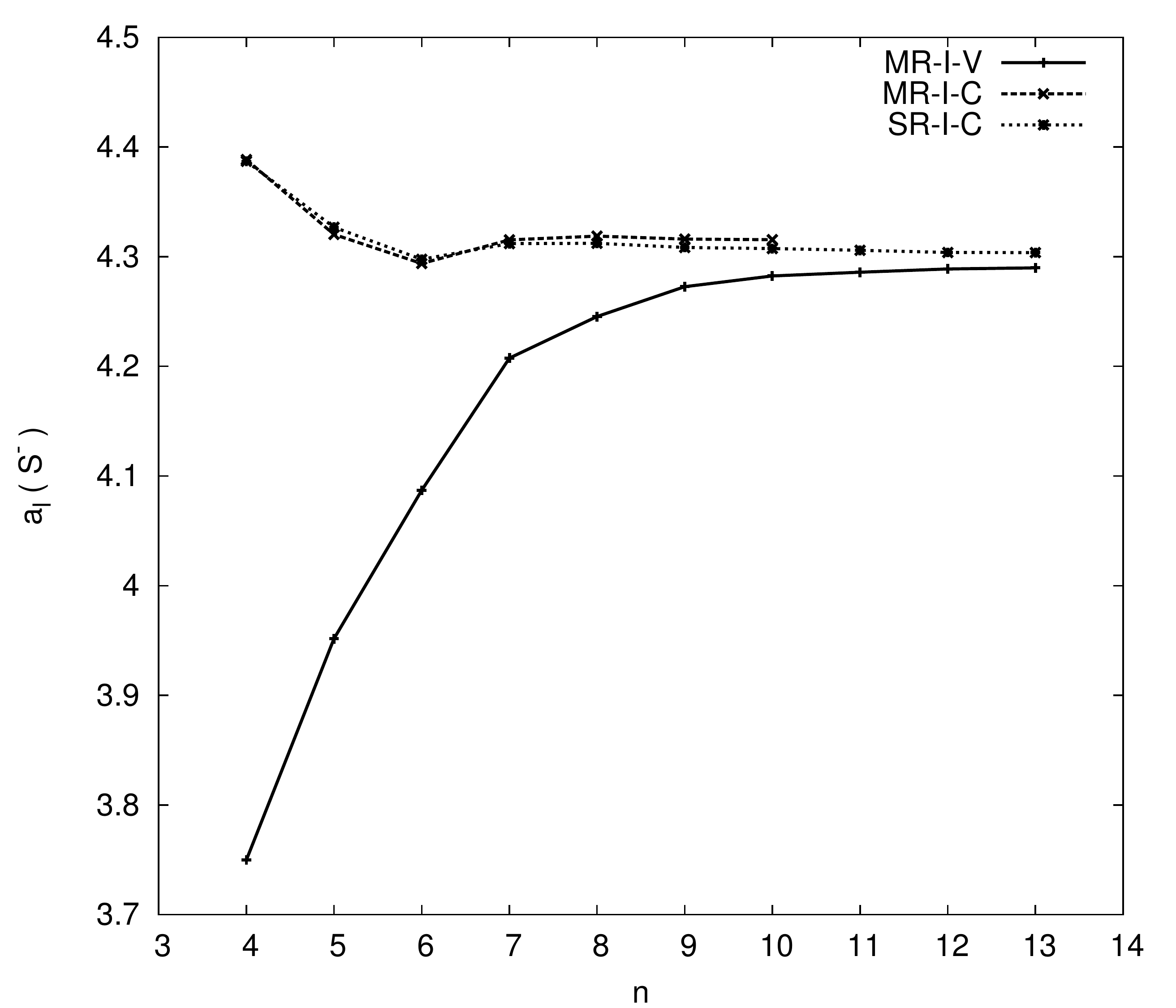}
\includegraphics[width=0.47\textwidth]{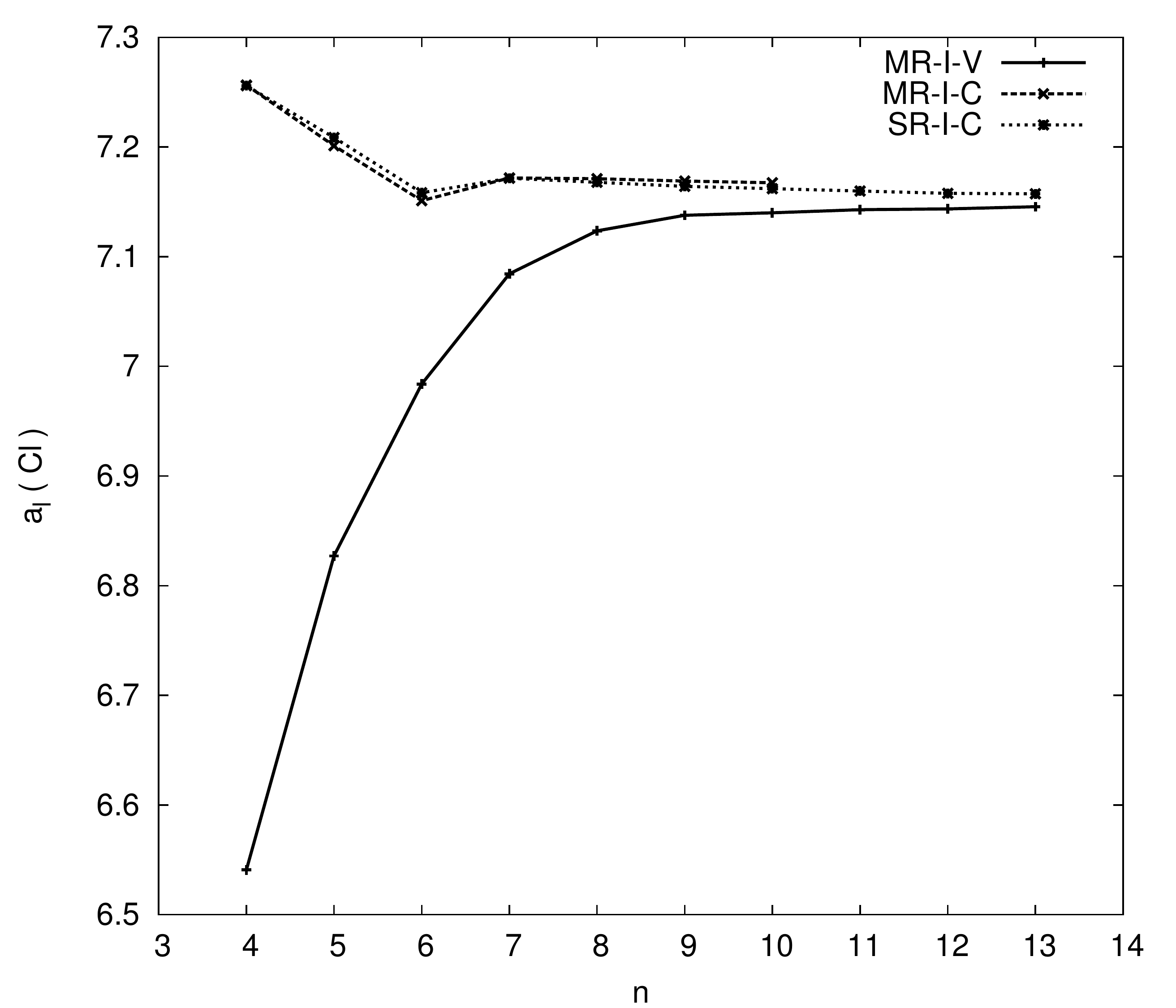}\\
\includegraphics[width=0.47\textwidth]{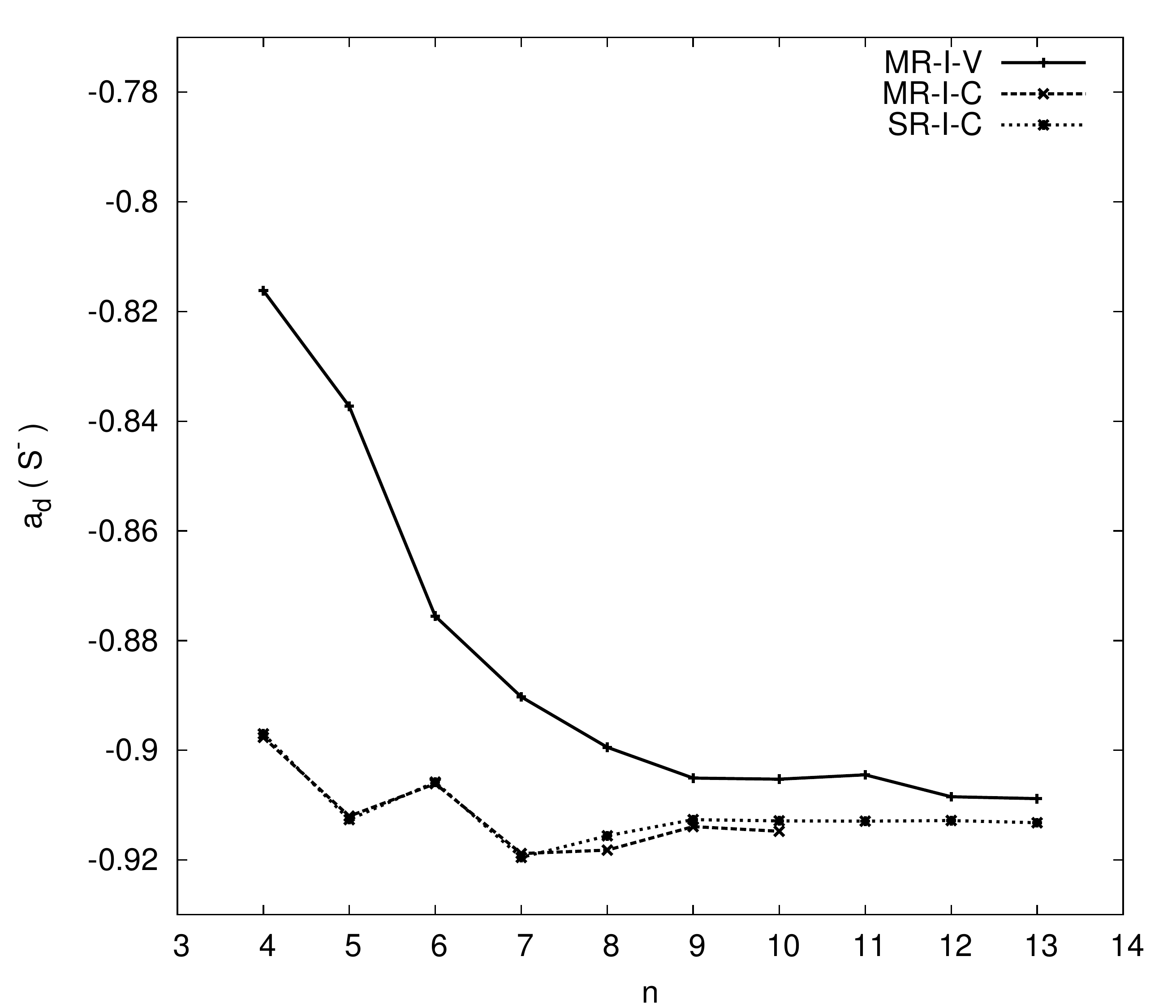}
\includegraphics[width=0.47\textwidth]{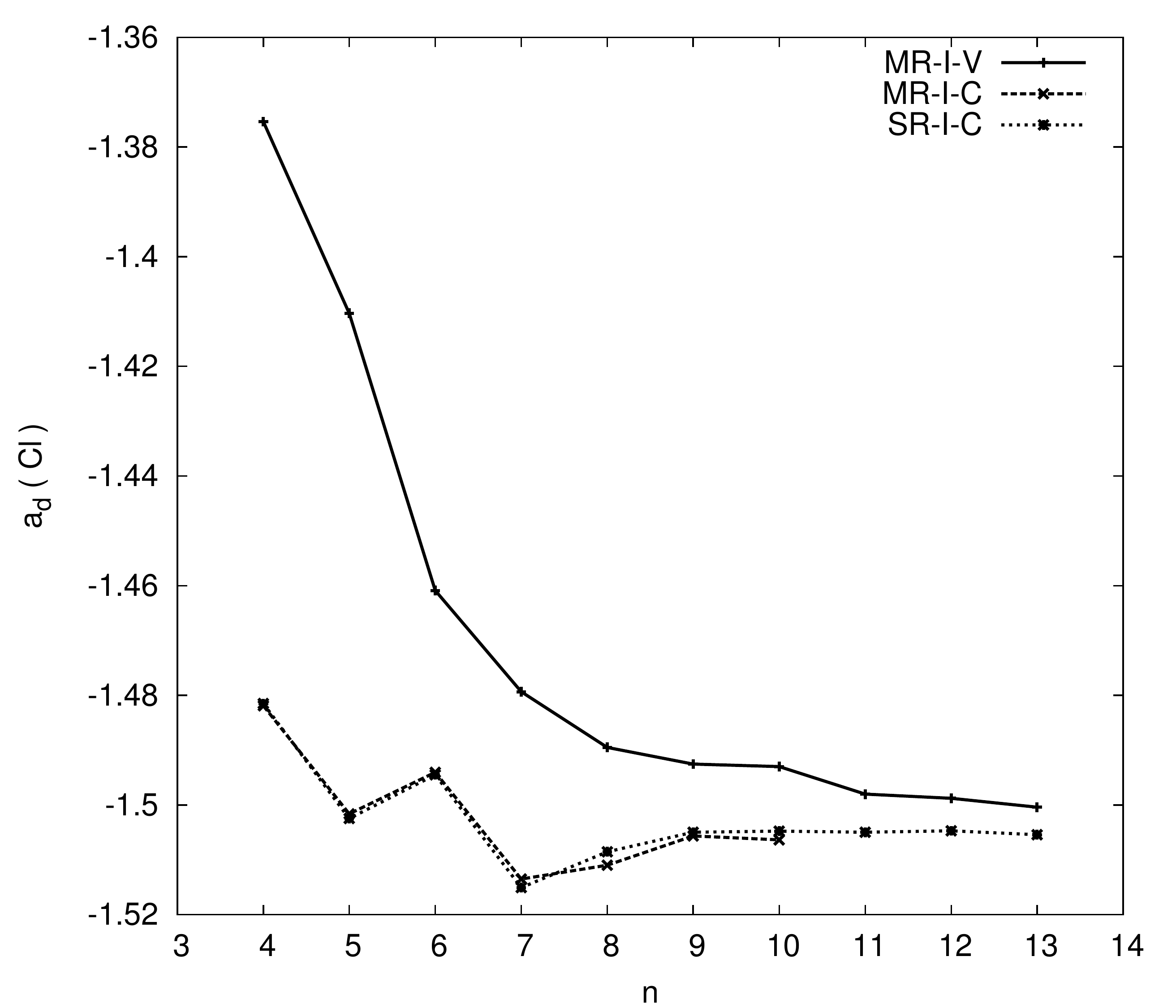}\\
\includegraphics[width=0.47\textwidth]{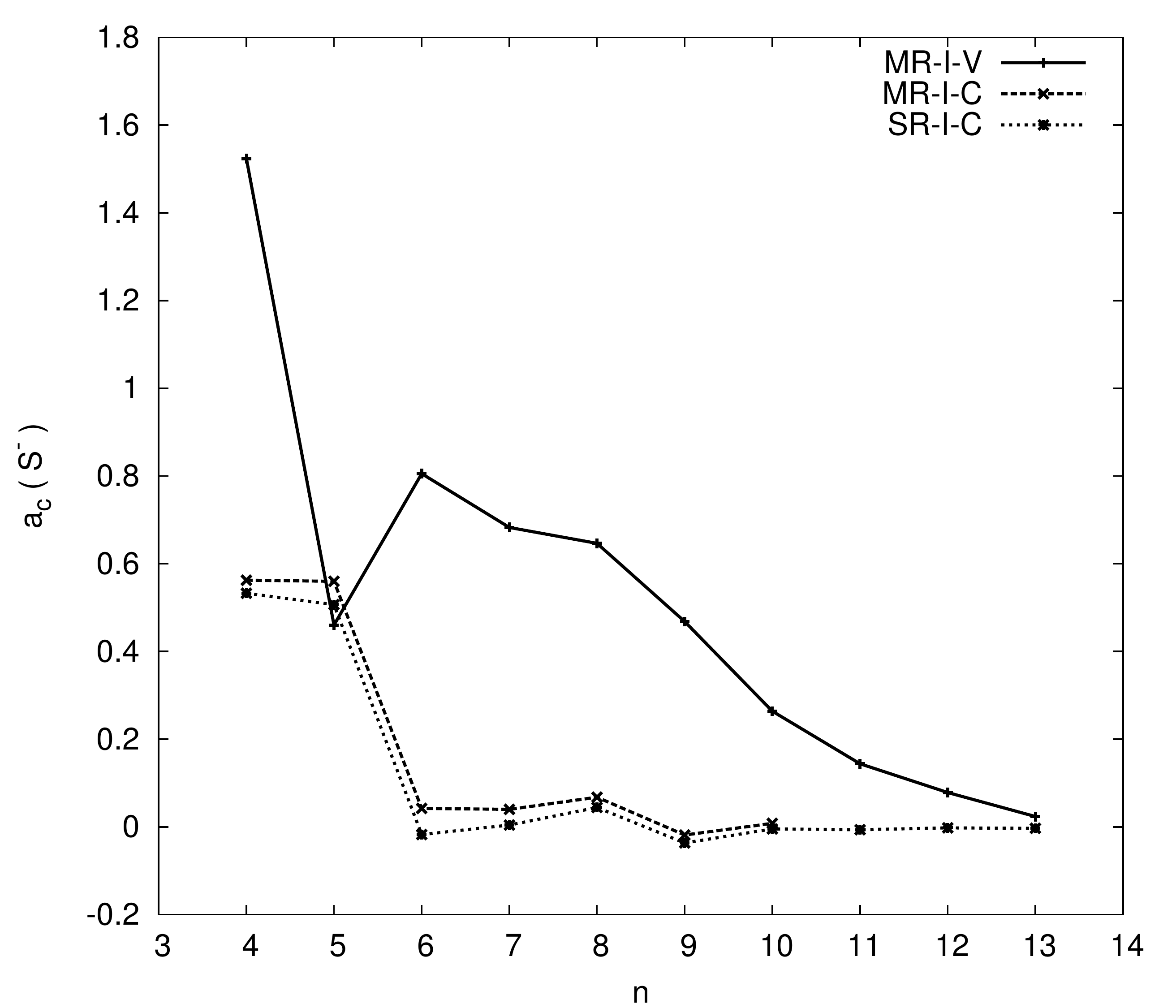}
\includegraphics[width=0.47\textwidth]{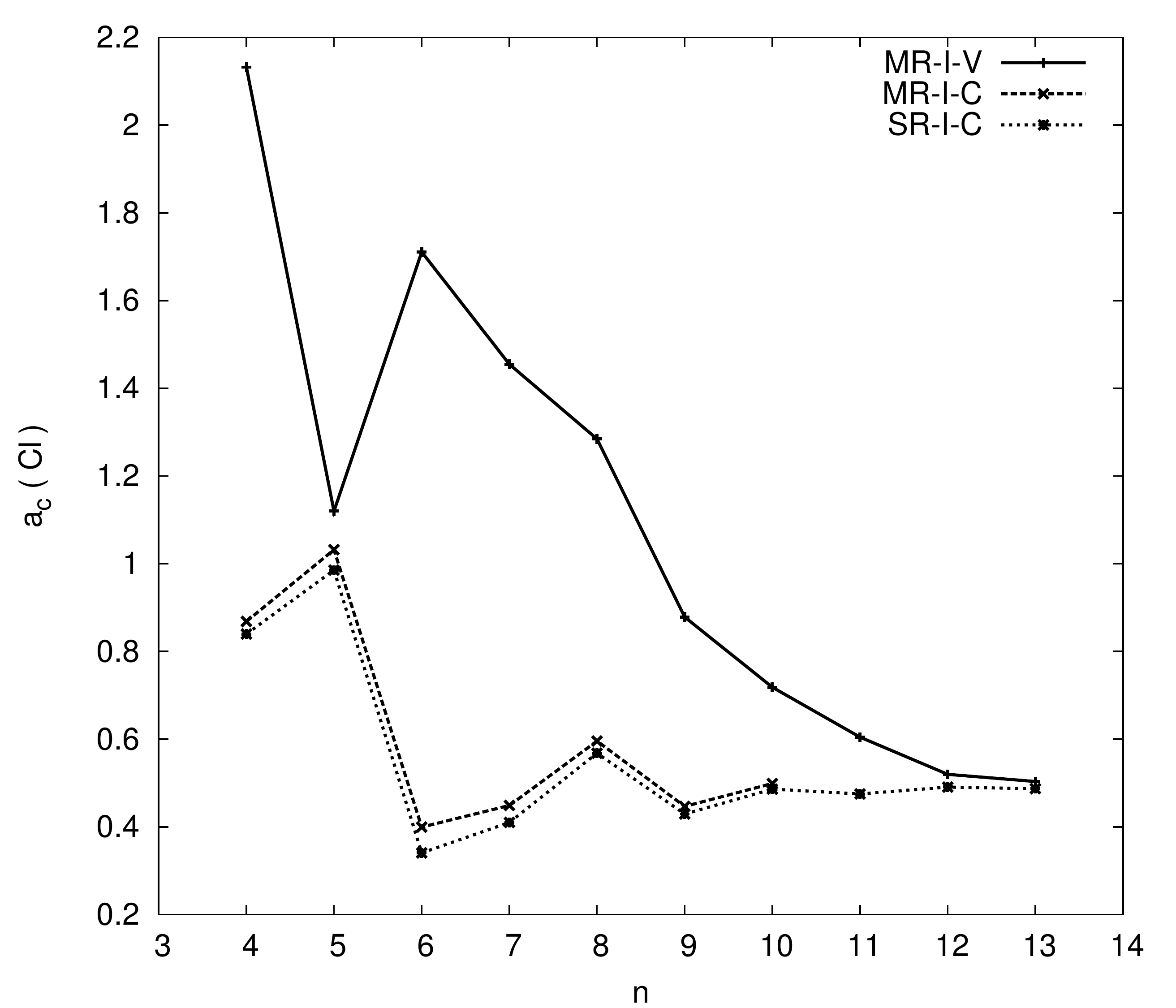}\\
\end{figure}

\begin{figure}
\caption
{For both S$^-$ (left plots) and Cl (right plots), values of non-relativistic electric quadrupole hyperfine interaction parameter $b$ (in $a_0^{-3}$) and energy $E$ (in E$_h$) obtained by CI calculations in the MR$_5$-I-C model with the three explored orbital basis sets SR-I-C, MR-I-C and MR-I-V, as a function of the active space $\lceil nl\rceil$, $l\leq 5$.\label{fig5:convb}}
\center
\includegraphics[width=0.47\textwidth]{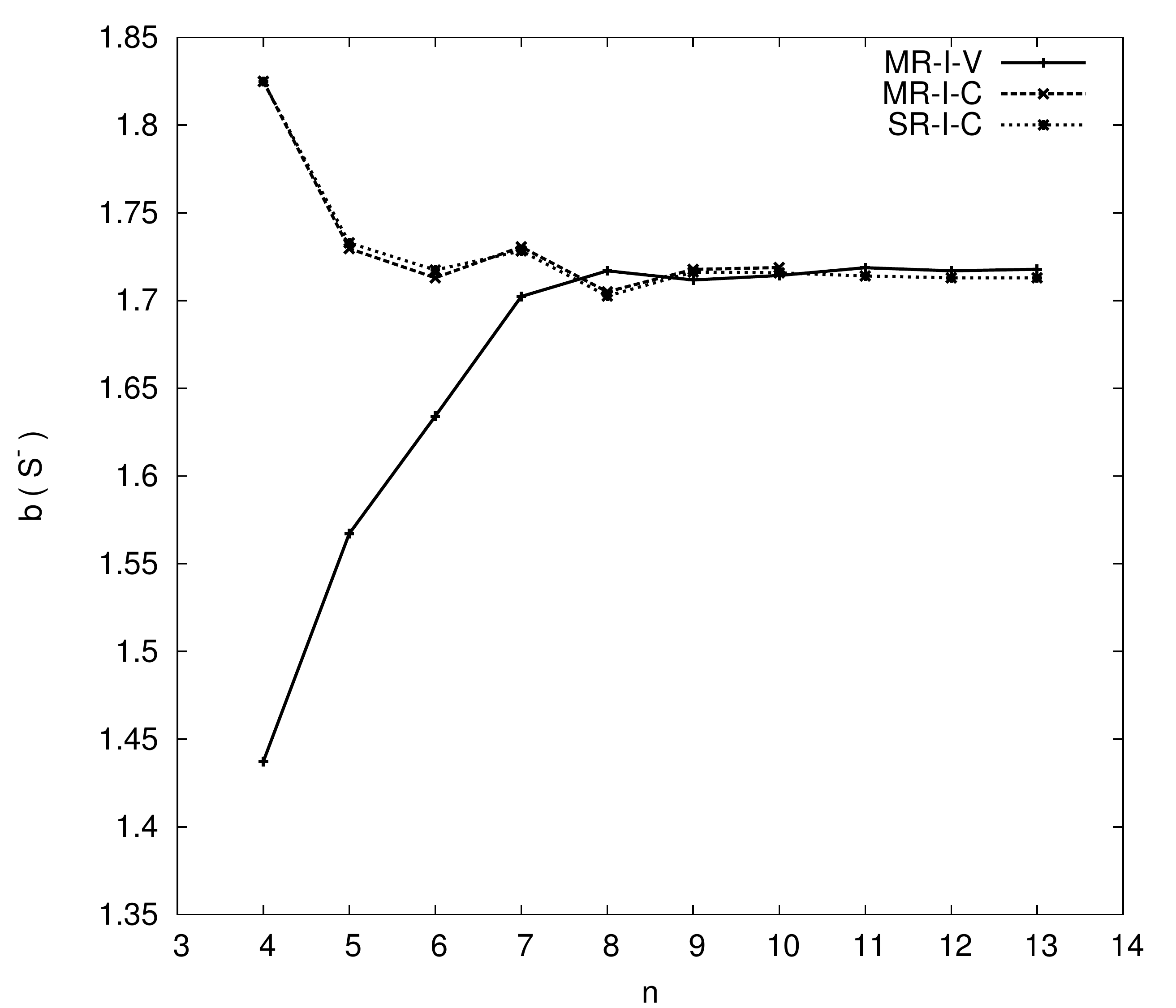}
\includegraphics[width=0.47\textwidth]{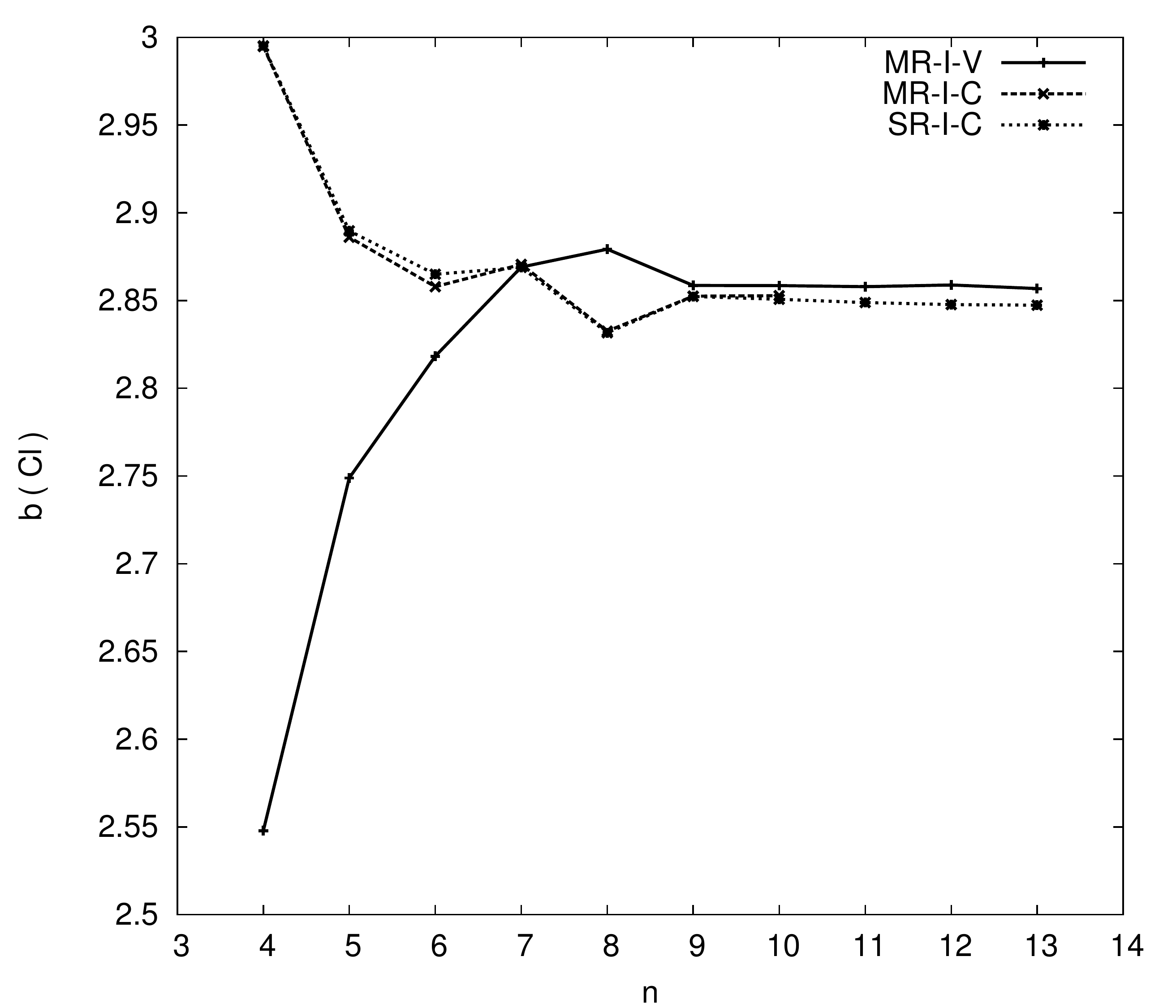}\\
\includegraphics[width=0.47\textwidth]{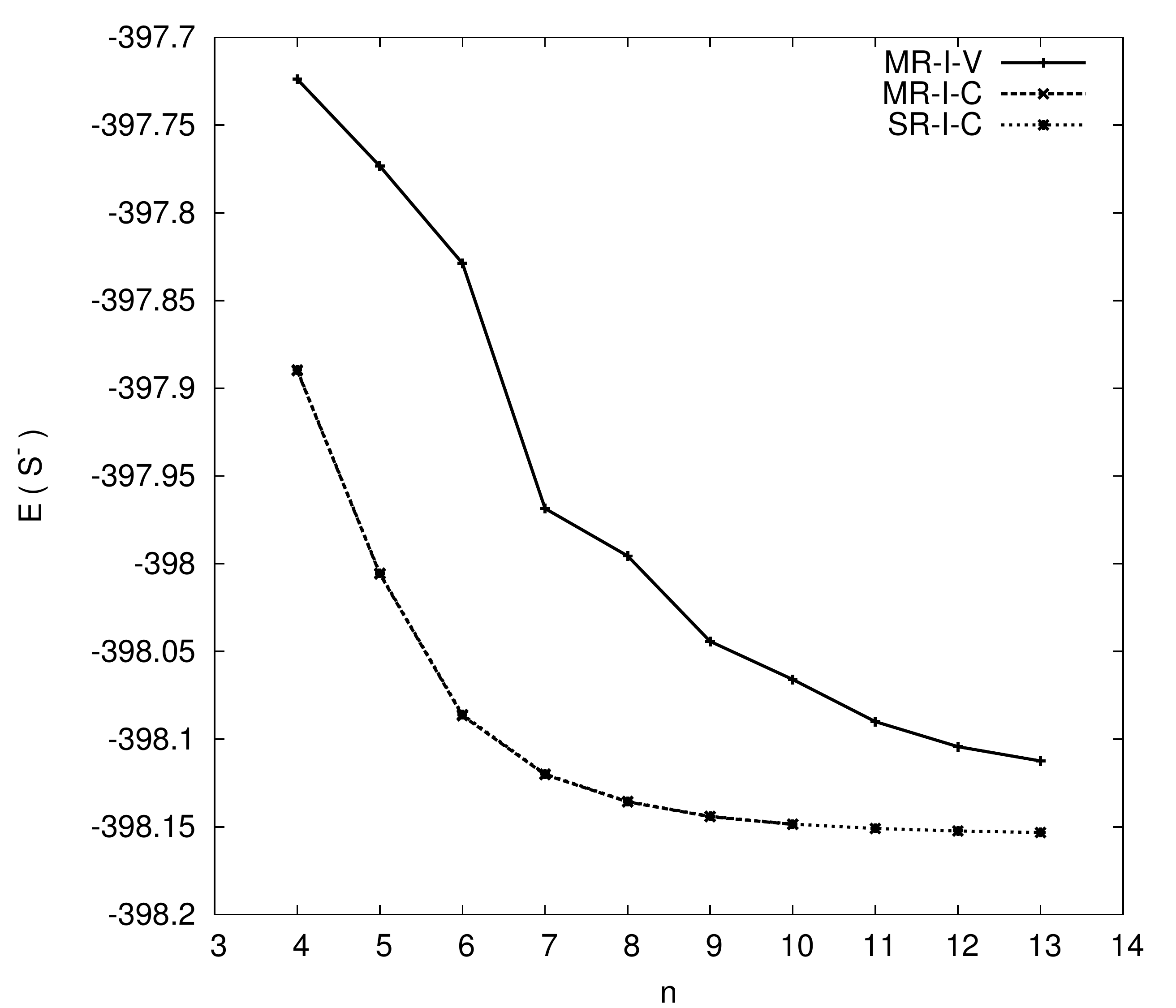}
\includegraphics[width=0.47\textwidth]{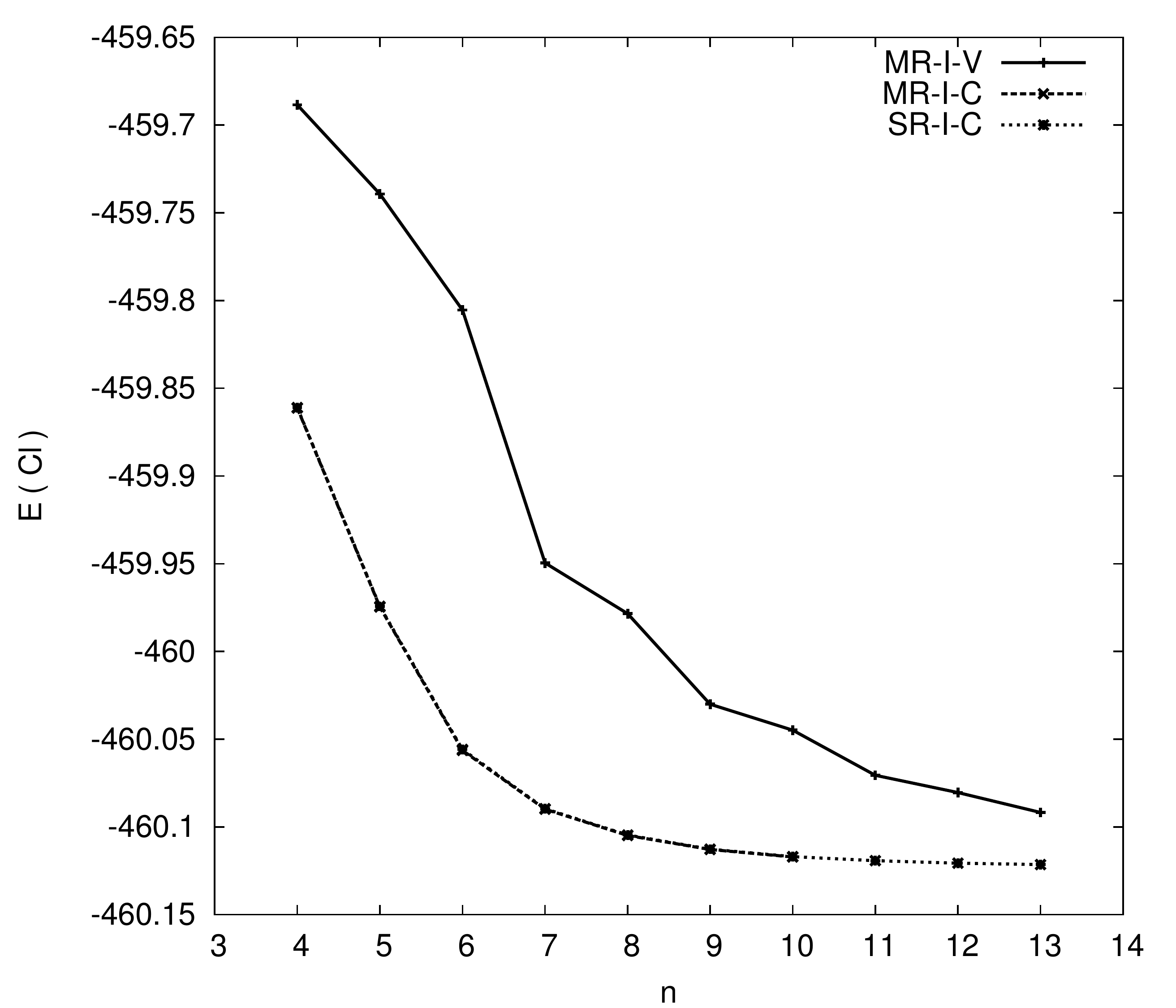}
\end{figure}

Figure~\ref{fig5:conva} and the two top plots of Figure~\ref{fig5:convb} present the convergence of the $a_l$, $a_d$, $a_c$ and $b$ hyperfine parameters (\ref{eq3:small_a_l}-\ref{eq3:small_b_hfs_par}) calculated with  the \mbox{MR$_5$-I-C} model in the three orbital basis sets (SR-I-C, MR-I-C and MR-I-V). All parameters are given in atomic units~($a_0^{-3}$). In all cases the MR-I-V$\lceil 13h \rceil$ and SR-I-C$\lceil 13h \rceil$ results differ by about $\lesssim 2\ 10^{-2}~a_0^{-3}$, the final $A_J$ constants themselves differing by less than half a percent. The convergence of the calculations with the orbital sets obtained from the MR-I-V model is slow compared to the one of the calculations based on the SR- and MR-I-C orbital sets.
It is remarkable to note that, to the contrary of the magnetic dipole parameters ($a$), the electric quadrupole parameter ($b$) converges as rapidly in the SR-I-C and MR-I-V basis sets.

Comparing the S$^-$ and Cl trends, it is the similarities that first strike. However, the $y$-axis scales being the same between the left (S$^-$) and corresponding right (Cl) plots, we observe easily that (i) the S$^-$ hyperfine parameters are shifted to smaller values compared to the Cl parameters, and (ii) the S$^-$ hyperfine structure calculations converge slightly faster that the Cl ones. These two observations can be easily understood by the mere diffuseness of the negative ion electron charge distribution. Indeed, when an electron attaches to a neutral atom, the valence shells spread and the core-valence separation becomes larger. Then, even though the negative ion have more electrons, the core-valence overlap is only slightly larger in the negative ion than in the neutral, even if there are more valence electrons in the former than in the latter~\cite{CarGod:11a,Car:10a}. Hence, the negative ion core is more spherical and the hyperfine constants tend to be smaller than in comparable systems. Consequently, even if negative ions are highly correlated systems, they are characterized by a comparatively smaller core-valence correlation. In particular, the hyperfine structure of negative ions is slightly less sensitive to correlation effects than one would first expect.


Even though the MR$_5$ and MR$_9$ sets do not account for exactly the same correlation effects depending on the active set, the approach based on the orbitals obtained at the SR-I-C$\lceil 13h \rceil$ level are superior. 
Tables~\ref{tab5:hfsparCl} and~\ref{tab5:hfsparS-} show the total energies and $a_l, a_d, a_c$ and $b$ parameters deduced from the MR$_5$-, MR$_9$-I-C$\lceil 12h \rceil$ and MR$_5$-I-C$\lceil 13h \rceil$ models using the SR-I-C orbitals. The final results are obtained by reporting the impact of the $13^{\textrm{th}}$ shell on the MR$_5$-I-C model on the MR$_9$ based values.

\begin{table}
\caption{Best estimates of the energy (in E$_h$) and hyperfine parameters (in $a_0^{-3}$) of Cl (see text). The final values are the results of the MR$_9$-I-C$\lceil 12h \rceil$ in which the impact of the $13^\textrm{th}$ correlation shell on the MR$_5$-I-C model is reported. \label{tab5:hfsparCl}}
\begin{indented}
\item[]
\begin{tabular}{cD{.}{.}{5}D{.}{.}{5}D{.}{.}{5}D{.}{.}{5}}
\br
& \multicolumn{2}{c}{MR$_5$}& \multicolumn{1}{c}{MR$_9$}\\ 
&\crule{2}&\crule{1}\\
& \multicolumn{1}{c}{$12h$} & \multicolumn{1}{c}{$13h$} & \multicolumn{1}{c}{$12h$} & \multicolumn{1}{c}{final}\\\mr
$E$  & -460.12061 & -460.12145 & -460.12315\\
$a_l$&     7.15768 &    7.15724 &    7.14014 &  7.13971\\
$a_d$&    -1.50471 &   -1.50541 &   -1.50369 & -1.50439\\
$a_c$&     0.49079 &    0.48733 &    0.52870 &  0.52525\\
$b$  &     2.84768 &    2.84738 &    2.83912 &  2.83882\\
\br
\end{tabular}
\end{indented}
\end{table}
\begin{table}
\caption{Best estimates of the energy (in E$_h$) and hyperfine parameters (in $a_0^{-3}$) of S$^-$ (see text). The final values are the results of the MR$_9$-I-C$\lceil 12h \rceil$ in which the impact of the $13^\textrm{th}$ correlation shell on the MR$_5$-I-C model is reported. \label{tab5:hfsparS-}}
\begin{indented}
\item[]
\begin{tabular}{cD{.}{.}{5}D{.}{.}{5}D{.}{.}{5}D{.}{.}{5}}
\br
& \multicolumn{2}{c}{MR$_5$}& \multicolumn{1}{c}{MR$_9$}\\ 
&\crule{2}&\crule{1}\\
& \multicolumn{1}{c}{$12h$} & \multicolumn{1}{c}{$13h$} & \multicolumn{1}{c}{$12h$} & \multicolumn{1}{c}{final}\\
\mr
$E$  & -398.15235 & -398.15324 & -398.15553\\
$a_l$&    4.30388 &    4.30366 &    4.28953 & 4.28931\\
$a_d$&   -0.91284 &   -0.91324 &   -0.91211 &-0.91251\\
$a_c$&   -0.00209 &   -0.00306 &    0.03581 & 0.03484\\
$b$  &    1.71294 &    1.71293 &    1.70529 & 1.70528\\
\br
\end{tabular}
\end{indented}
\end{table}

\section{Hyperfine Structures of the $3p^4~^3P_J$ levels of S}\label{sec5:S_HFS}

\subsection{Definition of the orbital set: MCHF calculations}

For the $3p^4~^3P$ state of neutral sulfur, we first select the references from which the MCHF expansions are generated.
Similarly to Cl and S$^-$, we use the mono-reference (SR) and a multireference
\beq
\textrm{MR}=\{1,2\}^{10} \{3s,3p\}^{4}\{3\}^2.
\eeq
The weights of the most important configurations in the SR-I-C, MR-I-C and MR-I-V calculations using the active set $\lceil 10h \rceil$ are presented in Table~\ref{tab5:weightsS}. The corresponding mean radius of the orbitals are shown in the fifth and sixth columns of Table~\ref{tab5:rad}. We observe that the eigenvector composition differences between the open-core and valence calculations are much more pronounced than in the S$^-$ and Cl cases. For instance, the $1s^22s^22p^43s^23p^44p^2$ configuration gains a significant weight in both open-core calculations, the $4p$ orbital being localized in the inner region of the atom. This important core effect is accompanied by a decrease of the closed-core configuration $3s^23p^2np^2$, $n=4$ for \mbox{MR-I-V} and $n=5$ for SR- and \mbox{MR-I-C} models.
In the open-core models, the cumulative weight of these two configurations is $0.055$, which is comparable to the weight of the $3s^23p^24p^2$ in the closed core model ($=0.068$). It indicates that the correlation of the $3p^4$ and $2p^6$ electrons is similar so that, even if we fix the core to its HF shape, high-order correlation effects mix the core and valence electrons through the $2p^43p^44p^2$ configuration, $(3p,4p)$ being variational. Similar observations were made in the context of the calculation of the electron affinity of Chlorine~\cite{CarGod:11a}, in which case the negative ion neon-like core can strongly mix with the valence $3p^6$ electrons compared to the neutral MCHF solution.

\begin{table}
\caption{Sorted weights of the first configurations in the SR-I-C, MR-I-C and MR-I-V wave functions of S. The active set is $\lceil 10h \rceil$. The $1s$ shell is closed in all those configurations.\label{tab5:weightsS}}
\vspace{1mm}
\begin{indented}
\item[]
\begin{tabular}{cD{.}{.}{4}cD{.}{.}{4}cD{.}{.}{4}}
\br
\multicolumn{2}{c}{SR-I-C} &  \multicolumn{2}{c}{MR-I-C}  &\multicolumn{2}{c}{MR-I-V} \\
\crule{2}&\crule{2}&\crule{2}\\
 config. & \multicolumn{1}{c}{$w$} & config. & \multicolumn{1}{c}{$w$} & config. & \multicolumn{1}{c}{$w$}\\
\mr
 $2s^2 2p^6 3s^2 3p^4          $ & 0.9726 & $2s^2 2p^6 3s^2 3p^4          $&0.9587&
  $3s^2 3p^4           $ & 0.9566 \\
 $2s^2 2p^6 3s^2 3p^2 3d^2     $ & 0.1139 & $2s^2 2p^6 3s^1 3p^4 3d^1     $&0.1574&
  $3s^1 3p^4 3d^1      $ & 0.1607 \\
 $2s^2 2p^6 3s^1 3p^4 3d^1     $ & 0.1106 & $2s^2 2p^6 3s^2 3p^2 3d^2     $&0.1547&
  $3s^2 3p^2 3d^2      $ & 0.1569 \\
 $2s^2 2p^6 3s^1 3p^3 3d^1 4f^1$ & 0.0419 & $2s^2 2p^6      3p^4 3d^2     $&0.0497&
  $3s^2 3p^2 4p^2      $ & 0.0677 \\
 $2s^2 2p^4 3s^2 3p^4 4p^2     $ & 0.0394 & $2s^2 2p^6 3s^1 3p^3 3d^1 4f^1$&0.0436&
  $3s^1 3p^3 4s^1 4p^1 $ & 0.0580 \\
 $2s^2 2p^6 3s^1 3p^3 3d^1 5p^1$ & 0.0394 & $2s^2 2p^6 3s^1 3p^3 3d^1 5p^1$&0.0393&
  $3s^1 3p^3 3d^1 4f^1 $ & 0.0562 \\
 $2s^2 2p^6      3p^4 3d^2     $ & 0.0389 & $2s^2 2p^4 3s^2 3p^4 4p^2     $&0.0389&
  $3s^1 3p^3 3d^1 4p^1 $ & 0.0547 \\
 $2s^2 2p^6 3s^2 3p^2 5p^2     $ & 0.0386 & $2s^2 2p^6 3s^2 3p^2 5p^2     $&0.0387&
  $     3p^4 3d^2      $ & 0.0498 \\
 $2s^2 2p^6 3s^1 3p^3 5s^1 5p^1$ & 0.0338 & $2s^2 2p^6 3s^1 3p^3 5s^1 5p^1$&0.0339&
  $3s^2 3p^3 4f^1      $ & 0.0489 \\
 $2s^2 2p^6 3s^2 3p^2 4p^1 5p^1$ & 0.0324 & $2s^2 2p^6 3s^1 3p^3 3d^1 6f^1$&0.0328&
  $3s^2 3p^3 4p^1      $ & 0.0486 \\
 $2s^2 2p^6 3s^1 3p^3 3d^1 6f^1$ & 0.0312 & $2s^2 2p^6 3s^2 3p^2 4p^1 5p^1$&0.0324&
  $3s^1 3p^3 3d^1 5f^1 $ & 0.0356 \\
 $2s^2 2p^4 3s^2 3p^4 4d^1 5d^1$ & 0.0289 & $2s^2 2p^4 3s^2 3p^4 4d^1 5d^1$&0.0285&
  $3s^2 3p^3 5p^1      $ & 0.0313 \\
 $2s^2 2p^6 3s^2 3p^3 4f^1     $ & 0.0271 & $2s^2 2p^6 3s^2 3p^3 4f^1     $&0.0281&
  $     3p^5 4p^1      $ & 0.0292 \\
 $2s^2 2p^4 3s^2 3p^4 5d^2     $ & 0.0260 & $2s^2 2p^4 3s^2 3p^4 5d^2     $&0.0256&
  $3s^1 3p^2 3d^3      $ & 0.0282 \\
 $2s^2 2p^4 3s^2 3p^4 4d^2     $ & 0.0260 & $2s^2 2p^4 3s^2 3p^4 4d^2     $&0.0256&
  $3s^2 3p^2 4f^2      $ & 0.0252 \\
 $2s^12p^53s^23p^44s^14p^1     $ & 0.0253 & $2s^12p^53s^23p^44s^14p^1     $&0.0250&
  $3s^13p^34p^14d^1    $ & 0.0247\\
\br
\end{tabular}
\end{indented}
\end{table}

\subsection{Fully correlated CI calculations}

\begin{table}
\caption{Best estimates of the energy (in E$_h$) and hyperfine parameters (in $a_0^{-3}$) of the neutral sulfur $^3P$ state. The final results are obtained by reporting the impacts of the $13^\textrm{th}$ correlation shell on the MR$_5$-I-C model and the impacts of the extension of the MR$_6$ reference to the MR$_8$ reference using the active set $\lceil 9h \rceil$, on the values of hyperfine parameters obtained in the MR$_6$-I-C$\lceil 12h \rceil$ calculation. \label{tab5:hfsparp4}}
\footnotesize
\begin{tabular*}{\textwidth}{cD{.}{.}{5}D{.}{.}{5}D{.}{.}{5}D{.}{.}{5}D{.}{.}{5}D{.}{.}{5}D{.}{.}{5}}
\br
& \multicolumn{3}{c}{MR$_5$}& \multicolumn{2}{c}{MR$_6$}& \multicolumn{1}{c}{MR$_8$} & \\ 
&\crule{3}&\crule{2}&\crule{1}\\
& \multicolumn{1}{c}{$9h$}& \multicolumn{1}{c}{$12h$} & \multicolumn{1}{c}{$13h$} & \multicolumn{1}{c}{$9h$} & \multicolumn{1}{c}{$12h$} & \multicolumn{1}{c}{$9h$}& \multicolumn{1}{c}{final}\\
\mr
$E$  & -398.07547 & -398.08254 & -398.08329 & -398.07655 & -398.08363 & -398.07952 & \\
$a_l$&    5.13968 &    5.12937 &    5.12869 &    5.14143 &    5.13020 &    5.13748 & 5.12557\\
$a_d$&    1.08020 &    1.07860 &    1.07918 &    1.08205 &    1.08039 &    1.08447 & 1.08339\\
$a_c$&    0.74528 &    0.85674 &    0.85825 &    0.80130 &    0.91908 &    0.82046 & 0.93976\\
$b$  &   -2.05860 &   -2.05035 &   -2.04957 &   -2.05942 &   -2.05064 &   -2.05571 &-2.04615\\
\br
\end{tabular*}
\end{table}

The large differences between the roles played by each correlation orbital in the open-core and valence calculations prevent us from comparing the two models as in the cases of S$^-$ and Cl.
Although the results of the different open-core CI models are coherent, we limit our discussion to the SR- and MR-I-C results.

Driven by the eigenvector composition analysis presented in Table~\ref{tab5:weightsS}, we choose the multi-reference
\beq
\textrm{MR}_5=\{ 3s^2 3p^4, 3s^23p^23d^2, 3s^1 3p^4 3d^1, 3s^1 3p^3 3d^1 4f^1, 3p^4 3d^2 \}.
\eeq
Then, for evaluating the impact of the addition of a core-polarization configuration in the reference, we build the MR
\beq
\textrm{MR}_{6} = \textrm{MR}_5 \cup \{ 2s^12p^53s^23p^44s^14p^1 \} \, ,
\eeq
and finally we define
\beq
\textrm{MR}_{8} = \textrm{MR}_{6} \cup \{ 2s^22p^43s^23p^44p^2, 2s^22p^63s^13p^33d^15p^1 \}.
\eeq
We show in Table~\ref{tab5:hfsparp4} the hyperfine parameters for the sulfur $3p^4~^3P$ ground state using these MR$_p$-I-C models ($p=5,6,8$) for the largest possible active set and using the \mbox{SR-I-C} radial functions. The final results are obtained by reporting the impacts of the $13^\textrm{th}$ correlation shell on the MR$_5$-I-C model and the impacts of the extension of the MR$_6$ reference to the MR$_8$ reference using the active set $\lceil 9h \rceil$, on the hyperfine parameter values obtained in the MR$_6$-I-C$\lceil 12h \rceil$ calculation.

\section{Comparison with experiment}\label{sec5:resHFS}

The three considered isotopes, $^{33}$S, $^{35}$Cl and  $^{37}$Cl, have a spin $I=3/2$ and respectively a magnetic dipole moment of $+0.6438212(14)~\mu_B$, $+0.8218743(4)~\mu_B$ and $+0.6841236(4)~\mu_B$~\cite{Sto:05a}. Their nuclear quadrupole moments are still best determined using theoretical inputs~\cite{Pyy:08a}, as we will see below.

The non-relativistic $A_J$ constants computed using the final set of hyperfine parameters of Tables~\ref{tab5:hfsparCl} , \ref{tab5:hfsparS-} and~\ref{tab5:hfsparp4} are shown in Table~\ref{tab5:AJ}.
We estimate the relativistic corrections by running mono-reference non-relativistic and corresponding relativistic CI calculations using the Pauli approximation with the \mbox{SR-I-C$\lceil 9h\rceil$} orbital set~\cite{CarGod:11b,Jonetal:10a}.

The neutral sulfur $A_1(^3P)$ hyperfine constant is characterized by strong cancellation between the spin-dipole ($A_{d}$) and orbit ($A_l$) contributions. Indeed, in the MR$_6$-I-C$\lceil 12h \rceil$ CI calculation performed with the SR-I-C orbitals, we find
\beq
A_l=105.04\textrm{ MHz} \qquad A_d=-110.74\textrm{ MHz} \qquad A_c=6.28\textrm{ MHz}\ .
\eeq
We realize from Table~\ref{tab5:hfsparCl}, \ref{tab5:hfsparS-} and \ref{tab5:hfsparp4} that the contact term ($A_c$) is by far the less converged contribution, bringing the largest source of uncertainty ($\sim$ 1 MHz). It is unclear to which extent the $^{33}$S$^-$ theory-experiment excellent agreement is accidental.

\begin{table}
\caption{Comparison of theoretical and experimental $A_J$ hyperfine constants (in MHz) for the lowest multiplet of $^{33}$S, $^{33}$S$^-$, $^{35}$Cl and $^{37}$Cl. Non-relativistic estimations (NR) computed from the final results of Tables~\ref{tab5:hfsparCl}, ~\ref{tab5:hfsparS-} and~\ref{tab5:hfsparp4}. Relativistic corrections estimated with a CI-RCI approach~\cite{CarGod:11b,Jonetal:10a}. \label{tab5:AJ}}
\footnotesize
\begin{tabular*}{\textwidth}{lD{.}{.}{2}D{.}{.}{2}D{.}{.}{2}D{.}{.}{2}D{.}{.}{2}D{.}{.}{2}D{.}{.}{2}D{.}{.}{2}}
\br
& \multicolumn{2}{c}{$^{33}$S}& \multicolumn{2}{c}{$^{33}$S$^-$}& \multicolumn{2}{c}{$^{35}$Cl}& \multicolumn{2}{c}{$^{37}$Cl}\\
&\crule{2}&\crule{2}&\crule{2}&\crule{2}\\
&\multicolumn{1}{c}{$A_{2}$}&\multicolumn{1}{c}{$A_{1}$}&\multicolumn{1}{c}{$A_{3/2}$}&\multicolumn{1}{c}{$A_{1/2}$}&\multicolumn{1}{c}{$A_{3/2}$}&\multicolumn{1}{c}{$A_{1/2}$}&\multicolumn{1}{c}{$A_{3/2}$}&\multicolumn{1}{c}{$A_{1/2}$}\\
\mr
Non-relativistic  & 133.58 & 0.33 & 92.48 & 483.30 & 168.52 & 846.09 & 202.45 & 1016.46\\
+rel. corrections & 130.09 & -4.62 & 91.43 & 496.28  & 167.52 & 872.05 & 201.25 & 1113.22\\
\\
Experiment & & & 91.49(9)^a & & 170.69^b & & 205.05^b & \\
\br
\end{tabular*}
$^a$ Reference \cite{Traetal:89a}.\\
$^b$ Reference \cite{Ful:76a}.
\end{table}

\begin{table}
\caption{Nuclear electric quadrupole moments ($Q$, in barns) estimated with equation~(\ref{eq5:Qp5}) for $^{33}$S$^-$, $^{35}$Cl and $^{37}$Cl. From those $Q$, we calculate the $B_J$ constants of $^{33}$S lower multiplet with our value of $b($S$~^3P)$. Non-relativistic estimations (NR) computed from the final values of Tables~\ref{tab5:hfsparCl}, ~\ref{tab5:hfsparS-} and~\ref{tab5:hfsparp4}. Relativistic corrections estimated with a CI-RCI approach~\cite{CarGod:11b,Jonetal:10a}.\label{tab5:Q}}
\footnotesize
\begin{tabular*}{\textwidth}{lD{.}{.}{8}D{.}{.}{2}D{.}{.}{2}D{.}{.}{8}D{.}{.}{8}}
\br
& \multicolumn{3}{c}{$^{33}$S}& \multicolumn{1}{c}{$^{35}$Cl}& \multicolumn{1}{c}{$^{37}$Cl}\\
&\crule{3}&\crule{1}&\crule{1}\\
&\multicolumn{1}{c}{$Q$}&\multicolumn{1}{c}{$B_2$($^{33}$S)\hspace*{0.3cm}}&\multicolumn{1}{c}{$B_1$($^{33}$S)\hspace*{0.3cm}}&\multicolumn{1}{c}{$Q$}&\multicolumn{1}{c}{$Q$}\\
\mr
Non-relativistic   & -0.0655(6) & -31.49 & 15.74 & -0.082265 & -0.064833 \\
+rel. corrections     &  -0.0657(6) & -31.60 & 15.80  &  -0.081764 & -0.064438\\
\\
Sundholm and Olsen$^a$ & -0.0678(13) & -32.60 & 16.30 & -0.08165(80) & -0.06435(64)\\
\br
\end{tabular*}
$^a$ Reference \cite{SunOls:90a} for $^{33}$S and \cite{SunOls:93a} for the chlorine isotopes
\end{table}

As far as the $B_J$ constants are concerned, it is more relevant to tabulate the electric quadrupole moments ($Q$) obtained for S$^-$ and Cl from the formula~\cite{Hib:75a}
\beq
Q=-\frac{(B_{3/2})_{exp}}{(b)_{th} G_q}
\label{eq5:Qp5}
\eeq
where $G_q=234.96475$ for obtaining $Q$ in barns when $b$ is in $a_0^{-3}$ and $B_J$ in MHz.
The available experimental data are:
\beqa
B_{3/2}(^{33}\textrm{S}^-) &=& 26.24(23)\phantom{0000~}~\textrm{MHz\qquad \cite{Traetal:89a}}\\
B_{3/2}(^{35}\textrm{Cl})  &=& 54.872~905(55)\textrm{ MHz\qquad \cite{Ful:76a}}\\
B_{3/2}(^{37}\textrm{Cl})  &=& 43.245~245(55)\textrm{ MHz\qquad \cite{Ful:76a}}
\eeqa
We compare our $Q$ values with previous works~\cite{SunOls:90a,SunOls:93a} in Table~\ref{tab5:Q}. Since, to our knowledge, there is no measurement of the neutral sulfur hyperfine structure, we use our value for the $b(\textrm{S}~^3P)$ parameter for estimating the $B_J$ constants of $^{33}$S with each $Q(^{33}\textrm{S})$ value.

\section{Conclusion}\label{sec5:concl}

We perform MCHF-CI and RCI calculations of the hyperfine constants of the $3p^5~^2P_J^o$ multiplet of $^{33}$S$^-$ and $^{35,37}$Cl and the $3p^4~^3P_J$ multiplet of $^{33}$S. We obtain good agreement with previous theoretical works~\cite{SunOls:90a,SunOls:93a} for the nuclear electric quadrupole moments of $^{33}$S and $^{37,35}$Cl, and with the $A(3p^5~^2P^o_{3/2})$ experimental values~\cite{Traetal:89a,Ful:76a}.
It appears that the contact contribution, the main source of uncertainty in our non-relativistic calculations, is ten times smaller in S$^-$ than in Cl. We interpret this as an effect of an increased separation of the core and valence regions in negative ions.

We show that, for sufficiently large active sets, orbitals optimized in closed-core MCHF calculations reproduce the results of proper open-core MCHF calculations. This approach has a significant advantage: the core-valence distinction in frozen and closed-core MCHF calculations is much cleaner. It allows to minimize the high-order core and valence mixing and hence get a better comparison between calculations performed on different systems (\eg S and S$^-$).

\section*{References}

\begin{thebibliography}{10}

\bibitem{Traetal:89a}
R~Trainham, R~M Jopson, and D~J Larson.
\newblock {\em Phys. Rev. A}, 39(7):3223, 1989.

\bibitem{SunOls:90a}
D Sundholm and J Olsen.
\newblock {\em Phys. Rev. A}, 42(3):1164, 1990.

\bibitem{Beretal:95a}
U~Berzinsh, M~Gustafsson, D~Hanstorp, A~Klinkm\"uller, U~Ljungblad, and
  A~M M{\aa}rtensson-Pendrill.
\newblock {\em Phys. Rev. A}, 51(1):231, 1995.

\bibitem{GodFro:99a}
M~R Godefroid and C {Froese Fischer}.
\newblock {\em Phys. Rev. A}, 60(4):R2640, 1999.

\bibitem{Bloetal:01a}
C~Blondel, C~Delsart, C~Valli, S~Yiou, M~R Godefroid, and S~Van~Eck.
\newblock {\em Phys. Rev. A}, 64(5):052{}504, 2001.

\bibitem{Caretal:10a}
T~Carette, C~Drag, O~Scharf, C~Blondel, C~Delsart, C~{Froese Fischer}, and M~R
  Godefroid.
\newblock {\em Phys. Rev. A}, 81(4):042{}522, 2010.

\bibitem{CarGod:11a}
T~Carette and M~Godefroid.
\newblock {submitted to J. Phys. B: At. Mol. Opt. Phys.}, 2011.

\bibitem{MadNov:74a}
D~L Mader and R~Novick.
\newblock {\em Phys. Rev. Lett.}, 32(5):185--188, 1974.

\bibitem{Fisetal:10a}
A~Fischer, C~Canali, U~Warring, A~Kellerbauer, and S~Fritzsche.
\newblock {\em Phys. Rev. Lett.}, 104(7):073{}004, 2010.

\bibitem{Veretal:10a}
S~Verdebout, P~J\"onsson, G~Gaigalas, M~Godefroid, and C~{Froese Fischer}.
\newblock {\em J. Phys. B: At. Mol. Opt. Phys.}, 43(7):074017, 2010.

\bibitem{Froetal:97a}
C~{Froese Fischer}, T~Brage, and P~J\"onsson.
\newblock {\em Computational Atomic Structure : An MCHF Approach}.
\newblock Taylor \& Francis, Inc., 1st edition, 1997.

\bibitem{Froetal:07a}
C~{Froese Fischer}, G~Tachiev, G~Gaigalas, and M~R Godefroid.
\newblock {\em Comp. Phys. Com.}, 176(8):559, 2007.

\bibitem{LinRos:74a}
I Lindgren and A Ros\'en.
\newblock {\em Case Stud. Atom. Phys.}, 4(3 and 4), 1974.

\bibitem{Hib:75a}
A~Hibbert.
\newblock {\em Rep. Prog. Phys.}, 38(11):1217, 1975.

\bibitem{Jonetal:93a}
P J\"onsson, C-G Wahlstr\"om, and C~{Froese Fischer}.
\newblock {\em Comp. Phys. Com.}, 74(3):399, 1993.

\bibitem{FroSax:74a}
C~{Froese Fischer} and K~M~S Saxena.
\newblock {\em Phys. Rev. A}, 9(4):1498, 1974.

\bibitem{LinMor:86a}
I Lindgren and J Morrison.
\newblock {\em Atomic Many-Body Theory (2nd. edition)}.
\newblock Number~3 in Springer Series on Atoms and Plasmas. Springer-Verlag,
  1986.

\bibitem{MigKim:98a}
J~Migdalek and Y~K Kim.
\newblock {\em J. Phys. B: At. Mol. Opt. Phys.}, 31(9):1947, 1998.

\bibitem{Fro:77a}
C~{Froese Fischer}.
\newblock {\em Hartree--Fock method for atoms. A numerical approach}.
\newblock John Wiley and Sons, Inc.,New York, 1977.

\bibitem{Godetal:98a}
M~R Godefroid, P J\"onsson, and C~{Froese Fischer}.
\newblock Atomic structure variational calculations in spectroscopy.
\newblock {\em Phys. Scr.}, T78:33, 1998.

\bibitem{Car:10a}
T Carette.
\newblock {\em Isotope effects in atomic spectroscopy of negative ions and
  neutral atoms: a theoretical contribution}.
\newblock PhD thesis, Universit\'e Libre de Bruxelles, 2010.
\newblock
  \url{http://theses.ulb.ac.be/ETD-db/collection/available/ULBetd-12132010-195%
442/}.

\bibitem{Sto:05a}
N~J Stone.
\newblock {\em At. Data Nucl. Data Tables}, 90(1):75, 2005.

\bibitem{Pyy:08a}
P Pyykk\"o.
\newblock {\em Mol. Phys.}, 106(16-18):1965, 2008.

\bibitem{CarGod:11b}
T Carette and M Godefroid.
\newblock {submitted to Phys. Rev. A}, 2011; arXiv:1101.5318v1

\bibitem{Jonetal:10a}
P~J{\"o}nsson, T~Carette, M~Nemouchi, and M~Godefroid.
\newblock {\em J. Phys. B: At. Mol. Opt. Phys.}, 43(11):115{}006, 2010.

\bibitem{Ful:76a}
G~H Fuller.
\newblock {\em J. Phys Chem. Ref. Data}, 5(4):835, 1976.

\bibitem{SunOls:93a}
D Sundholm and J Olsen.
\newblock {\em J. Chem. Phys.}, 98(9):7152, 1993.

\end{thebibliography}

\end{document}